\documentclass[aps,twocolumn,superscriptaddress]{revtex4}

\usepackage{graphicx}
\usepackage{epsfig}             

\begin{document}


\title{Continuous-Variable Spatial Entanglement for Bright Optical
Beams}

\author{Magnus T.L. Hsu}
\affiliation{Australian Centre for Quantum-Atom Optics, Department
of Physics, Australian National University, ACT 0200, Australia}

\author{Warwick P. Bowen} \altaffiliation{now at the Quantum Optics
Group, MC 12-33, Norman Bridge Laboratory of Physics, California
Institute of Technology, Pasadena CA 91125.} \affiliation{Australian
Centre for Quantum-Atom Optics, Department of Physics, Australian
National University, ACT 0200, Australia}

\author{Nicolas Treps}
\affiliation{Laboratoire Kastler Brossel, Universit\'e Pierre et
Marie Curie, case 74, 75252 Paris cedex 05, France}

\author{Ping Koy Lam}
\email[Email: ]{ping.lam@anu.edu.au} \affiliation{Australian
Centre for Quantum-Atom Optics, Department of Physics, Australian
National University, ACT 0200, Australia}

\date{\today}

\begin{abstract}
A light beam is said to be position squeezed if its position can be
determined to an accuracy beyond the standard quantum limit.  We
identify the position and momentum observables for bright optical
beams and show that position and momentum entanglement can be
generated by interfering two position, or momentum, squeezed beams on
a beam splitter.  The position and momentum measurements of these
beams can be performed using a homodyne detector with local oscillator
of an appropriate transverse beam profile.  We compare this form of
spatial entanglement with split detection-based spatial entanglement.
\end{abstract}

\pacs{42.50, 42.30}

\maketitle

\section{Introduction}

The concept of entanglement was first proposed by Einstein, Podolsky
and Rosen in a seminal paper in 1935 \cite{EPR}. The original
Einstein-Podolsky-Rosen (EPR) entanglement, as discussed in the paper,
involved the position and momentum of a pair of particles.  In this
article, we draw an analogy between the original EPR entanglement and
the position and momentum ($x$-$p$) entanglement of bright optical
beams.

Entanglement has been reported in various manifestations.
For continuous wave (CW) optical beams, these
include, quadrature \cite{silberhorn, ou} and polarisation
\cite{bowen-pol} entanglement. Spatial forms of entanglement,
although well studied in the single photon regime, have not been
studied significantly in the continuous wave regime. Such forms of
entanglement are interesting as they span a potentially
infinite Hilbert space. Spatial EPR entanglement \cite{lugiato}
has wide-ranging applications from two-photon quantum imaging
\cite{abouraddy, pittman} to holographic teleportation
\cite{abouraddy1, sokolov} and interferometric faint phase object
quantum imaging \cite{sokolov1}.

Current studies are focused on $x$-$p$ entanglement for the few
photons regime. Howell {\it et al.} \cite{howell} observed near
and far-field quantum correlation, corresponding to the position
and momentum observables of photon pairs. Gatti {\it et al.}
\cite{gatti4} have also discussed the spatial EPR aspects in the
photons pairs emitted from an optical parametric oscillator below
threshold. Other forms of spatial entanglement which are related to
image correlation have also been investigated. A scheme to produce
spatially entangled images between the signal and idler fields
from an optical parametric amplifier has been proposed by Gatti
{\it et al.} \cite{gatti, gatti1}. Their work was extended to the
macroscopic domain by observing the spatial correlation between
the detected signal and idler intensities, generated via the
parametric down conversion process \cite{gatti3}.

Our proposal considers the possibility of entangling the position
and momentum of a free propagating beam of light, as opposed to
the entanglement of local areas of images, considered in previous
proposals. Our scheme is based on the concept of position squeezed
beams where we have shown that we have to squeeze the transverse
mode corresponding to the first order derivative of the mean field in
order to generate the position squeezed beam \cite{hsu}. Similarly
to the generation of quadrature entangled beams, the position
squeezed beams are combined on a beam-splitter to generate $x$-$p$
entangled beams. We also propose to generate spatial entanglement
for split detection, utilising spatial squeezed beams reported by
Treps {\it et al.} \cite{treps1d, treps2d, treps2dlong}. This form
of spatial entanglement has applications in quantum imaging
systems.

We first define the position and momentum of an optical beam by
performing a multi-modal decomposition on a displaced and tilted beam,
respectively.  We consider the case of a TEM$_{00}$ beam and show
that the corresponding position and momentum observables are conjugate
observables which obey the Heisenberg commutation relation.  We then
propose a scheme to produce $x$-$p$ entanglement for TEM$_{00}$
optical beams.  Finally, we consider spatial squeezed beams for split
detectors and show that it is also possible to generate spatial
entanglement with such beams.

\section{Position-Momentum Entanglement}

\subsection{Definitions - Classical Treatment}

\begin{figure}[!ht]
\begin{center}
\includegraphics[width=5.5cm]{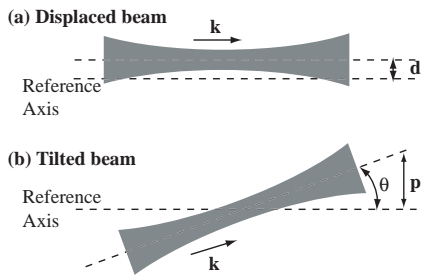}
\caption{(a) Beam displacement $d$, and (b) beam tilt by angle $\theta$, from
a reference axis.}
\label{disp-tilt}
\end{center}
\end{figure}
Let us consider an optical beam with $x$- and $y$- symmetric
transverse intensity profile propagating along the $z$-axis.  Since
the axes of symmetries remain well defined during propagation, we can
relate the beam position relative to these axes.  To simplify our
analysis we henceforth assume without loss of generality, a
one-dimensional beam displacement, $d$, from the reference $x$-axis
(see Fig.~\ref{disp-tilt}(a)).  We denote the electric field profile of
the beam by $E(x)$.  For a displaced beam, the electric field profile
is given by
\begin{equation} \label{udisp}
E_{d}(x) = E(x) + d \frac{\partial E(x)}{\partial x} +
\frac{d^2}{2} \frac{\partial^2 E(x) }{\partial x^2} + \cdots
\end{equation}
In the regime where displacement is much smaller than the beam
size, we can utilise the linearised approximation where only the
zeroth and first order terms are significant.  We see from
this expression that the zeroth order term is not dependent
on $d$, and that the displacement is directly proportional
to the derivative of the field amplitude $\partial E(x) /\partial
x$ \cite{hsu}.

The transverse beam momentum $p$ on the other hand, can be
obtained from the transverse component of the wave-number of the
beam, $p=k \sin \theta$, where $k=2\pi/\lambda$ and the beam
tilt is $\theta$.  This beam tilt is defined with respect to a
pivot point at the beam waist, as shown in
Fig.~\ref{disp-tilt}(b).

The electric field profile for a tilted beam with untilted
electric field profile $E(x)$ and wavelength $\lambda$ is given
by
\begin{equation} \label{utilt}
E_{\theta}(x) = \exp \left[ \frac{ i 2 \pi x \sin \theta }{
\lambda } \right] E(x \cos \theta)
\end{equation}
We can again simplify Eq.~(\ref{utilt}) by taking the zeroth and first
order Taylor expansion terms to get a transverse beam momentum of $p
\simeq k\theta$.  In the case of small displacement or tilt, we
therefore obtain a pair of equations
\begin{eqnarray}
E_{d}(x) &\approx E(x)  + \ d & \! \frac{\partial E(x)}{\partial x} \label{utilt1}\\
E_{p}(x) &\approx E(x)  + \ p & \!i x E(x) \label{utilt2}
\end{eqnarray}
Eqs.~(\ref{utilt1}) and (\ref{utilt2}) give the field parameters
that relate to the displacement and tilt of a beam.  For freely
propagating optical modes, the Fourier transform of the
derivative of the electric field, ${\cal F}(\partial E(x)/\partial
x)$, is of the form $ix {E}(x)$.  Hence, the Fourier transform of
displacement is tilt. 

In the case of a single photon, the position and momentum are
defined by considering the spatial probability density of the
photon, given by $|E(x)|^2/I$, where $I=\int|E(x)|^2dx$ is the 
normalisation factor. The mean position obtained from an 
ensemble of measurements on single
photons is then given by $\langle x \rangle = \frac{1}{I}\int
x|E(x)|^2dx$. The momentum of the photon is defined by the
spatial probability density of the photon in the far-field, or
equivalently by taking the Fourier transform of $\langle x
\rangle$. These definitions are consistent with our definitions of
position and momentum for bright optical modes.

\subsection{TEM$_{pq}$ Basis}

In theory, spatial entanglement can be generated for fields
with any arbitrary transverse mode-shape.  However, as with other
forms of continuous-variable entanglement, the efficacy of
protocols to generate entanglement is highest if the initial
states are minimum uncertainty.  For position and momentum
variables, the minimum uncertainty states are those which satisfy
the Heisenberg uncertainty relation $\Delta^{2} \hat{x} \Delta^{2}
\hat{p} \ge \hbar$, in the equality.  This equality is only
satisfied by states with Gaussian transverse distributions \cite{griffiths},
therefore we limit our analysis to that of TEM$_{00}$
modes.

A field of frequency $\omega$ can be represented by the
positive frequency part of the mean electric field
$\mathcal{E}^{+} e^{i\omega t}$. We are interested in the
transverse information of the beam described fully by the slowly
varying field envelope $\mathcal{E}^{+}$. We express this field in
terms of the TEM$_{pq}$ modes. For a measurement performed in an
exposure time $T$, the mean field for a displaced TEM$_{00}$ beam
can be written as
\begin{equation} \label{Ed}
\mathcal{E}_{d}^{+} (x) = i \sqrt{\frac{\hbar
\omega}{2\epsilon_{0} c T}} \sqrt{N} \Big(u_{0}(x) +
\frac{d}{w_{0}} u_{1}(x) \Big)
\end{equation}
where the first term indicates that the power of the displaced
beam is in the TEM$_{00}$ mode while the second term gives the
displacement signal contained in the  amplitude of the TEM$_{10}$
mode component. The corresponding mean field for a tilted
TEM$_{00}$ beam can be written as
\begin{equation} \label{Ep}
\mathcal{E}_{p}^{+} (x) = i \sqrt{\frac{\hbar
\omega}{2\epsilon_{0} c T}} \sqrt{N} \Big(u_{0}(x) + \frac{i
w_{0}p}{2} u_{1}(x) \Big)
\end{equation}
where the second term describes the beam momentum signal,
contained in the $\pi/2$ phase-shifted TEM$_{10}$ mode component.

\subsection{Definitions - Quantum Treatment}

We now introduce a quantum mechanical representation of the 
beam by taking into account the quantum noise of optical modes.
We can write the positive frequency part of the electric field
operator in terms of photon annihilation operators $\hat{a}$. The
field operator is given by
\begin{equation} \label{general}
\hat{\mathcal{E}}_{\rm in}^{+} = i \sqrt{\frac{\hbar \omega}{2
\epsilon_{0} c T}} \sum_{n=0}^{\infty} \hat{a}_{n} u_{n} (x)
\end{equation}
where $u_{n}(x)$ are the transverse beam amplitude functions for
the TEM$_{pq}$ modes and $\hat{a}_{n}$ are the corresponding annihilation operators. $\hat{a}_{n}$ is normally written in the form of $\hat{a}_{n}=
\langle \hat{a}_{n} \rangle +\delta \hat{a}_{n}$, where $\langle
\hat{a}_{n} \rangle$ describes the coherent amplitude part and
$\delta \hat{a}_{n}$ is the quantum noise operator.

In the small displacement and tilt regime, the TEM$_{00}$ and
TEM$_{10}$ modes are the most significant \cite{hsu}, with the
TEM$_{10}$ mode contributing to the displacement and tilt signals.
We can rewrite the electric field operator for mean number of
photons $N$ as
\begin{eqnarray}
\hat{\mathcal{E}}_{in}^{+} & = & i \sqrt{\frac{\hbar \omega}{2
\epsilon_{0} c T}} \Big(\sqrt{N} u_{0} (x) +
\frac{\delta\hat{X}_{a_{0}}^{+} + i\delta\hat{X}_{a_{0}}^{-}}{2}
\nonumber\\
 & & + \sum_{n = 1}^{\infty} \Big( \frac{\hat{X}_{a_{i}}^{+} + i
 \hat{X}_{a_{i}}^{-} }{2} \Big) u_{i} (x) \Big)
\end{eqnarray}
where the annihilation operator is now written in terms of
the amplitude $\hat{X}_{a}^{+}$ and phase $\hat{X}_{a}^{-}$
quadrature operators.

The displacement and tilt of a TEM$_{00}$ beam is given by the
amplitude and phase of the $u_{1}(x)$ mode in Eqs.~(\ref{Ed}) and
(\ref{Ep}), respectively.  We can, therefore, write the beam
position and momentum operators as
\begin{eqnarray}
\hat{x} &=& \frac{w_{0}}{2 \sqrt{N}} \hat{X}_{a_{1}}^{+} \label{xcoh}\\
\hat{p} &=& \frac{1}{ w_{0} \sqrt{N}} \hat{X}_{a_{1}}^{-} \label{pcoh}
\end{eqnarray}

\subsection{Commutation Relation}

Two observables corresponding to the position and momentum of a
TEM$_{00}$ beam have been defined.  We have shown that the position
and momentum observables correspond to near- and far-field
measurements of the beam, respectively.  Hence, we expect from
Eqs.~(\ref{xcoh}) and (\ref{pcoh}), that the position and momentum
observables do not commute.  Indeed, the commutation relation between
the two quadratures of the TEM$_{10}$ mode is
$[\hat{X}_{a_{1}}^{+},\hat{X}_{a_{1}}^{-}] = 2i$.  This leads to the
commutation relation between the position and momentum observables of
an optical beam with $N$ photons
\begin{equation} \label{comm}
[\hat{x},\hat{p} ] = \frac{i}{N}
\end{equation}
This commutation relation is similar to the position-momentum
commutation relation for a single photon, aside from the $1/N$
factor. The $1/N$ factor is related to the precision with which
one can measure beam position and momentum. Rewriting the
Heisenberg inequality using the commutation relation, gives
\begin{equation}
    \Delta^2\hat x\Delta^2\hat p\geq\frac{1}{4N}.
\end{equation}

The position measurement of a coherent optical beam gives a signal
which scales with $N$. The associated quantum noise of the
position measurement scales with $\sqrt{N}$. Hence the positioning
sensitivity of a coherent beam scales as $\sqrt{N}$ \cite{hsu,
treps1d}. The same consideration applied to the sensitivity of
beam momentum measurement shows an equivalent dependence of
$\sqrt{N}$. This validates the factor of $N$ in the Heisenberg
inequality and the commutation relation for a CW optical beam.

\subsection{Entanglement Scheme}

\begin{figure}[!ht]
\begin{center}
\includegraphics[width=8cm]{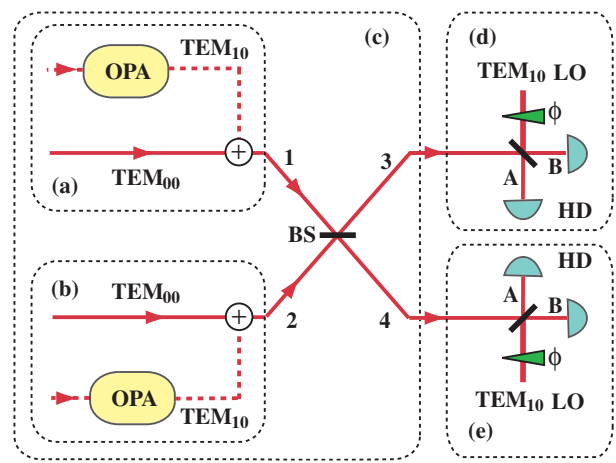}
\caption{Scheme for generating position-momentum entanglement for
continuous wave TEM$_{00}$ optical beams. OPA: optical parametric
amplifiers for the generation of squeezed light, BS: 50:50
beam-splitter, HD: homodyne detectors, LO: local oscillator beams,
$\phi$: phase of local oscillator beam.} \label{entg}
\end{center}
\end{figure}

We have shown that the position and momentum observables of CW
TEM$_{00}$ optical beams satisfy the Heisenberg commutation
relation. Consequently, EPR entanglement for the position and
momentum of TEM$_{00}$ beams is possible. Experimentally, the
usual quadrature entanglement is generated by mixing two amplitude
squeezed beams on a 50:50 beamsplitter. The same idea can be
applied to generate EPR $x$-$p$ entanglement, by using position
squeezed beams \cite{hsu}. Our scheme to produce $x$-$p$
entanglement between two CW TEM$_{00}$ optical beams is shown in
Figure~\ref{entg}.
The position squeezed beams in (a) and (b) are generated via the
in-phase combination of a vacuum squeezed TEM$_{10}$ beam with a
coherent TEM$_{00}$ beam. Such beam combination can be achieved
experimentally, for example using an optical cavity or a
beam-splitter \cite{treps2dlong}. The result of the combination is
a position squeezed beam. To generate entanglement, we consider beams with zero mean position and momentum, but we are interested in the quantum noise of the position and momentum of the beam. With this assumption, the electric field operators for the two input beams at the beam-splitter are given by
\begin{equation} \label{beam1}
\hat{\mathcal{E}}_{1}^{+} = i\sqrt{\frac{\hbar \omega}{2
\epsilon_{0} c T}} \Big( \sqrt{N} u_{0}(x) +  \sum_{n=0}^{\infty}
\delta \hat{a}_{n} u_{n}(x) \Big)
\end{equation}
\begin{equation} \label{beam2}
\hat{\mathcal{E}}_{2}^{+} = i\sqrt{\frac{\hbar \omega}{2
\epsilon_{0} cT}} \Big( \sqrt{N} u_{0}(x) + \sum_{n=0}^{\infty}
\delta \hat{b}_{n} u_{n}(x) \Big)
\end{equation}
where in both equations, the first bracketed term describes the
coherent amplitude of the TEM$_{00}$ beam. The second bracketed
terms describe the quantum fluctuations present in all modes. For position squeezed states, only the TEM$_{10}$ mode is occupied by a vacuum squeezed mode. All other modes are occupied by vacuum fluctuations. It is also assumed that the number of photons in each of the two beams, during the measurement window, is equal to $N$. The two position squeezed beams (1,2) are combined in-phase on a 50:50 beam-splitter (BS) in (c).

The usual input-output relations of a beam-splitter apply. The
electric field operators describing the two output fields from the
beam-splitter are given by $\hat{\mathcal{E}}_{3} =
(\hat{\mathcal{E}}_{1} + \hat{\mathcal{E}}_{2})/\sqrt{2}$ and
$\hat{\mathcal{E}}_{4} = (\hat{\mathcal{E}}_{1} -
\hat{\mathcal{E}}_2)/\sqrt{2}$. To demonstrate the existence of
entanglement, we seek for quantum correlation and anti-correlation
between the position and momentum quantum noise operators. The
position operator corresponding to beams 3 and 4 are given
respectively by
\begin{eqnarray}
\delta \hat{x}_{3} & = & \frac{w_{0} }{2\sqrt{2} \sqrt{N}} \Big( \delta \hat{X}_{a_{1}}^{+} + \delta \hat{X}_{b_{1}}^{-} \Big) \label{x3} \nonumber\\
 & = & \frac{1}{\sqrt{2}} \Big( \delta \hat{x}_{a} + \frac{w_{0} ^{2}}{2} \delta \hat{p}_{b} \Big)\\
\delta \hat{x}_{4} & =& \frac{w_{0} }{2\sqrt{2} \sqrt{N}} \Big( \delta \hat{X}_{a_{1}}^{+} - \delta \hat{X}_{b_{1}}^{-} \Big)  \nonumber\\
 & = & \frac{1}{\sqrt{2}} \Big( \delta \hat{x}_{a} - \frac{w_{0} ^{2}}{2} \delta \hat{p}_{b} \Big) \label{x4}
\end{eqnarray}
The momentum operator corresponding to the photo-current
difference for beams 3 and 4 are given by
\begin{eqnarray}
\delta \hat{p}_{3} & = & \frac{1}{w_{0} \sqrt{2 \sqrt{N}}} \Big( \delta \hat{X}_{a_{1}}^{-} + \delta \hat{X}_{b_{1}}^{+} \Big)  \nonumber\\
 & = & \frac{1}{\sqrt{2}} \Big( \delta \hat{p}_{a} + \frac{2}{w_{0}^{2}} \delta \hat{x}_{b} \Big)\label{p3}\\
\delta \hat{p}_{4} & = & \frac{1}{w_{0} \sqrt{2 \sqrt{N}}} \Big( \delta \hat{X}_{a_{1}}^{-} - \delta \hat{X}_{b_{1}}^{+} \Big) \nonumber\\
 & = & \frac{1}{\sqrt{2}} \Big( \delta \hat{p}_{a} - \frac{2}{w_{0}^{2}} \delta \hat{x}_{b} \Big) \label{p4}
\end{eqnarray}
In our case where the two input beams are position squeezed, the
sign difference between the position noise operators in
Eqs.~(\ref{x3}) and (\ref{x4}) as well as that between the
momentum noise operators in Eqs.~(\ref{p3}) and (\ref{p4}) are
signatures of correlation and anti-correlation for $\delta
\hat{x}$ and $\delta \hat{p}$.

\subsection{Inseparability Criterion}

Many criterions exist to characterise entanglement, for example
the \emph{inseparability criterion} \cite{duan} and the \emph{EPR
criterion} \cite{reid}. We have adopted the {\it inseparability
criterion} to characterise position-momentum entanglement. For
states with Gaussian noise statistics, Duan {\it et al.}
\cite{duan} have shown that the {\it inseparability criterion} is
a necessary and sufficient criterion for EPR entanglement.

In the case where two beams are perfectly interchangeable and have
symmetrical fluctuations in the amplitude and phase quadratures,
the {\it inseparability criterion} has been generalised and
normalised to a product form given by \cite{bowen-pol, bowen-pol1,
bowen-epr, bowen-epr1, tan, mancini}
\begin{equation}
\mathcal{I} (\hat{x}, \hat{p}) = \frac{ \langle (\hat{x}_{3} +
\hat{x}_{4})^{2} \rangle \langle (\hat{p}_{3} - \hat{p}_{4})^{2}
\rangle }{ | [\hat{x}, \hat{p}] |^{2}}
\end{equation}
for any pair of conjugate observables $\hat{x}$ and $\hat{p}$,
and a pair of beams denoted by the subscripts 3 and 4. For
states which are inseparable, $\mathcal{I} (\delta \hat{x}, \delta
\hat{p}) < 1$. By using observables $\delta \hat{x}$ and $\delta
\hat{p}$ from Eqs.~(\ref{x3}), (\ref{x4}), (\ref{p3}) and
(\ref{p4}) as well as the commutation relation of Eq. (\ref{comm})
the inseparability criterion for beams 3 and 4 is given by
\begin{eqnarray}
\mathcal{I} (\delta \hat{x}, \delta \hat{p}) & = &
\frac{16N^2}{\omega_0^4} \langle (\delta \hat{x}_{a_{1}}^{+} )^{2}
\rangle \langle (\delta
\hat{x}_{b_{1}}^{+} )^{2} \rangle \nonumber\\
& = & \langle (\delta \hat{X}_{a_{1}}^{+} )^{2} \rangle \langle
(\delta \hat{X}_{b_{1}}^{+} )^{2} \rangle \nonumber \\
& < & 1
\end{eqnarray}
where we have assumed that the TEM$_{10}$ modes of beams 1 and 2 are amplitude squeezed (i.e. $\langle (\delta \hat{X}_{a_{1}}^{+})^{2} \rangle <
1$ and $\langle (\delta \hat{X}_{b_{1}}^{+})^{2} \rangle < 1$).

Thus we have demonstrated that CV EPR entanglement between the
position and momentum observables of two CW beams can be achieved.

\subsection{Detection Scheme}

Ref.~\cite{hsu} has shown that the optimum small displacement
measurement is homodyne detection with a TEM$_{10}$ local
oscillator beam (see Fig.~\ref{entg}~(d)). When the input beam is
centred with respect to the TEM$_{10}$ local oscillator beam, no
power is contained in the TEM$_{10}$ mode. Due to the
orthogonality of Hermite-Gauss modes, the TEM$_{10}$ local
oscillator only detects the TEM$_{10}$ vacuum noise component. As
the input beam is displaced, power is coupled into the TEM$_{10}$
mode. This coupled power interferes with the TEM$_{10}$ local
oscillator beam, causing a change in photo-current observed at the
output of the homodyne detector. Thus the difference photo-current
of the TEM$_{10}$ homodyne detector is given by \cite{hsu}
\begin{equation} \label{ndiffx}
\hat{n}_{-}^{d} = \frac{2 \sqrt{N}\sqrt{N_{\rm LO}} }{ w_{0}}
\hat{x}
\end{equation}
where $N_{\rm LO}$ and $N$ are the total number of photons in the
local oscillator  and displaced beams, respectively, with $N_{\rm
LO} \gg N$. The linearised approximation is utilised, where second
order terms in $\delta \hat{a}$ are neglected since $N \gg
|\langle \delta \hat{a}_{n}^{2} \rangle|$ for all $n$.

In order to measure momentum, one could use a lens to
Fourier transform to the far-field plane, where the beam is then measured using the TEM$_{10}$ homodyning scheme. However, we have shown that the
the position and momentum of a TEM$_{00}$ beam differs by the phase of the
TEM$_{10}$ mode component. Indeed for a tilted TEM$_{00}$ beam,
the TEM$_{10}$ mode component is $\pi/2$ phase shifted relative to
the TEM$_{00}$ mode component. Consequently the phase quadrature of the
TEM$_{10}$ mode has to be interrogated. This can be achieved by
utilising a TEM$_{10}$ local oscillator beam with a $\pi/2$ phase
difference relative to the TEM$_{10}$ mode component of the TEM$_{00}$ beam. The resulting photo-current difference between the two homodyning detectors, for $N_{\rm LO} \gg N$, is given by
\begin{equation}
\hat{n}_{-}^{p}  = w_{0} \sqrt{N} \sqrt{N_{\rm LO}} \hat{p}
\end{equation}
%

\section{Spatial entanglement for split detection}

The entanglement presented in the previous section is analogous to $x$-$p$ entanglement in the single photon regime. However, the choice of the mean field mode is restricted to the TEM$_{00}$ mode. This limits the richness of a spatial variable and thus excludes the possibility of generating an infinite Hilbert space. To exploit the properties of spatial variables, we now consider more traditional forms of spatial squeezing. Consequently, we study the possibility of generating spatial entanglement for array detection devices, based on spatial squeezed beams.

\subsection{Spatial Squeezing}

Spatial squeezing was first introduced by Kolobov \cite{Kolobov}.
The generation of spatial squeezed beams for split and array
detectors was experimentally demonstrated by Treps {\it et al.} \cite{treps2d,
treps2dlong, treps1d}. A one-dimensional spatial squeezed beam
has a spatially ordered distribution, where there exists
correlation between the photon numbers in both transverse halves
of the beam. A displacement signal applied to this beam can thus
be measured to beyond the QNL.

We consider a beam of normalised transverse amplitude function $v_{0}(x)$
incident on a split detector. The noise of split detection has been shown to be due to the flipped mode \cite{fabre}, given by
\begin{displaymath}
v_{1}(x) = \left\{ \begin{array}{ll}
 v_{0}(x) & \textrm{for } x > 0 \\
 - v_{0}(x) & \textrm{for } x < 0
 \end{array} \right.
\end{displaymath}

When the field is centred at the split-detector, such that
the mean value of the measurement is zero, the flipped mode is
thus orthogonal to the mean field mode. In this instance, modes
$v_{i}(x)$ (for $i>1$) can be derived to complete the modal basis.
The electric field operator written in this new modal basis is given by
\begin{equation} \label{SDin}
\hat{\mathcal{E}}^{+} = i\sqrt{\frac{\hbar \omega}{2 \epsilon_{0}
cT}} \Big( \sqrt{N} v_{0} (x) +\sum_{n=0}^{\infty} \delta
\hat{c}_{n} v_{n} (x) \Big)
\end{equation}
where the first term describes the coherent excitation of the beam
in the $v_{0}(x)$ mode and $N$ is the total number of photons in the
beam. It has been shown that the corresponding photon number difference operator for split detection is given by \cite{hsu}
\begin{equation} \label{SDamp}
\hat{n}_{-}^{(+)} = \sqrt{N} \delta \hat{X}_{c_{0}}^{+}
\end{equation}

The beam is spatially squeezed if the state of the flipped mode is
vacuum squeezed and in phase with the mean field mode (see Fig.~\ref{spentg}~(a) and (b)).

\subsection{Spatial Homodyne}

Since split detection is commonly used as a detection device for beam position, one would naturally consider taking the Fourier transform of a spatial squeezed beam to obtain the conjugate observable for the beam. However, we have shown that split detection does not correspond exactly to beam position measurement. Thus the Fourier plane of the spatial squeezed beam does not provide the conjugate observable. More practically, the flipped mode is not mode-shape invariant under Fourier transformation. In the far-field, each odd-ordered mode component of the flipped mode obtains a $(2n+1)\pi$ Gouy phase difference, compared to the near-field. Thus the mode-shape in the far-field is no longer a flipped mode. Consequently, far and near-field measurements of a
spatial squeezed beam will not give the conjugate observables.

However, we can find the conjugate observables of a spatial squeezed beam by drawing an analogy to standard homodyne detection. In split detection, the equivalent local oscillator mode is the mean field $v_{0}(x)$ mode. The mode under interrogation by the split detector is the flipped mode $v_{1}(x)$. In the case of homodyne detection, the phase of the local oscillator beam is varied to measure the conjugate observables (i.e. amplitude and phase quadratures) of the input beam. Adapting this concept to the split detector, the conjugate observables for the spatial squeezed beam is thus the amplitude and phase quadratures of the flipped mode, while the mode-shape of the flipped mode remains unaltered. This is further verified upon inspection of Eq.~(\ref{SDamp}).

Our scheme to perform a phase measurement of the flipped mode
is shown in Fig.~\ref{spentg}~(d). In our scheme we assume that the
mean field is a TEM$_{00}$ mode. Note that in principle, this analysis could be performed for any mode-shape. The coherent TEM$_{00}$ mode component provides a phase reference for the flipped mode, analogous to that of a local
oscillator beam in homodyne detection. Thus the phase quadrature of
the flipped mode can be accessed by applying a $\pi/2$ phase shift
between the the TEM$_{00}$ mode and the flipped mode noise
component. Experimentally, this is achievable using an optical
cavity. When the cavity is non-resonant for the $v_{0}(x)$ and
$v_{1}(x)$ modes it will reflect off the two modes, in phase, onto
the split detector. This will give a measurement of the amplitude
quadrature of the flipped mode. However, the cavity can be tuned
to be partially resonant on the $v_{0}(x)$ mode while reflecting
the flipped mode. A $\pi/2$ phase difference can then be introduced
between the reflected $v_{0}(x)$ and $v_{1}(x)$ modes, giving a
measurement of the phase quadrature of the flipped mode. The
corresponding photon number difference operator is
\begin{equation} \label{SDphase}
\hat{n}_{-}^{(-)} = \sqrt{N} \delta \hat{X}_{c_{0}}^{-}
\end{equation}
which is the orthogonal quadrature of the spatial squeezed beam. The photon number operators corresponding to the two measurements in Eqs.~(\ref{SDamp}) and (\ref{SDphase}) are conjugate observables and satisfy the commutation relation $[\hat{n}_{-}^{(+)}, \hat{n}_{-}^{(-)} ] = 2iN$.

It is important to realise that the number of photons $N$ in Eqs.~(\ref{SDamp}) and (\ref{SDphase}) are only approximately equal. This is due to the fact that partial power in the TEM$_{00}$ mode is transmitted by the cavity, when the cavity is partially resonant on the TEM$_{00}$ mode. Although it is possible to implement a scheme that conserves the total number of photons at detection (e.g. losslessly separating the mean field and flipped modes and recombining them with a phase difference), we would like to emphasise that our scheme is more simple and intuitive, as well as being valid when $N$ is large.

\subsection{Entanglement Scheme}

\begin{figure}[!ht]
\begin{center}
\includegraphics[width=7cm]{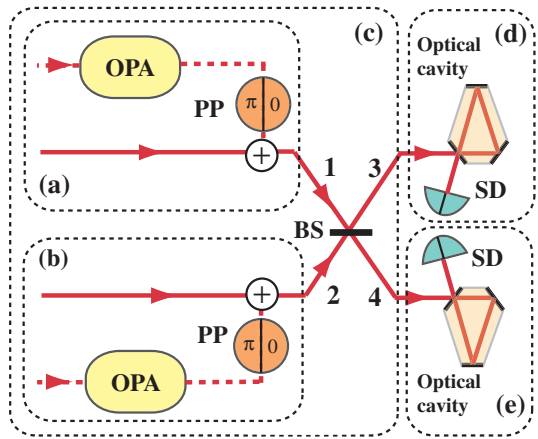}
\caption{Scheme for generating spatial entanglement for TEM$_{00}$
continuous wave light beams. OPA: optical parametric amplifiers
for generating squeezed beams, PP: phase plates for producing
flipped modes, BS: 50:50 beam-splitter.} \label{spentg}
\end{center}
\end{figure}

In order to generate spatial entanglement for split detection, two
spatial squeezed beams labelled 1 and 2 are combined on a 50:50
beam-splitter, as shown in Figure~\ref{spentg}~(c).

The electric field operators for the two input spatial squeezed
beams at the beam-splitter are described in a form identical to
that of Eq.~(\ref{SDin}). The annihilation operators of the
electric field operators for input beams 1 and 2 are labelled by
$\hat{c}_{n}$ and $\hat{d}_{n}$, respectively. By following a
similar procedure as before, the photon number difference operator
for output beams 3 and 4 from the beam-splitter are calculated.

For the amplitude quadrature measurement, the addition of the
difference photo-current between beams 3 and 4 yields
\begin{equation} \label{nsum}
\hat{n}_{3-}^{(+)} + \hat{n}_{4-}^{(+)} = \sqrt{N} ( \delta
\hat{X}_{c_{0}}^{+} +  \delta \hat{X}_{d_{0}}^{+})
\end{equation}
For the phase quadrature measurement, the subtraction of the
difference photo-current between beams 3 and 4 gives
\begin{equation} \label{ndiff}
\hat{n}_{3-}^{(-)} - \hat{n}_{4-}^{(-)} = \sqrt{N} ( \delta
\hat{X}_{d_{0}}^{+} -  \delta \hat{X}_{c_{0}}^{+})
\end{equation}

To verify spatial entanglement, the {\it inseparability criterion}
is utilised. The substitution of  Eqs.~(\ref{nsum}), (\ref{ndiff})
and the commutation relation between the photon number difference
operators into the generalised form of the {\it inseparability
criterion} gives
\begin{eqnarray}
\mathcal{I} (\delta \hat{n}_{-}^{(+)}, \delta \hat{n}_{-}^{(-)}) & = & \frac{ N \left( V_{c_{0}}^{2} + 2 V_{c_{0}}V_{d_{0}} + V_{d_{0}}^{2} \right) }{ 4N} \nonumber\\
& < & 1
\end{eqnarray}
where $V_{c_{0}} = \langle (\delta \hat{X}_{c_{0}}^{+})^{2}
\rangle$ and $V_{d_{0}} = \langle (\delta \hat{X}_{d_{0}}^{+})^{2}
\rangle$ are the variances for the flipped mode component of the
spatial squeezed beams 1 and 2. The inseparability criterion is
satisfied for amplitude squeezed flipped modes $V_{c_{0}} < 1$ and
$V_{d_{0}} < 1$.

We have proposed a scheme to generate spatial entanglement for split detection, using spatial squeezed beams. Spatial squeezing has been defined for any linear measurement performed with an array detector \cite{trepsmulti}. Similarly, spatial entanglement corresponding to any linear measurement, can be obtained. For an infinite span array detector with infinitessimally small pixels, it is thus possible to generate multi-mode spatial entanglement, increasing the Hilbert space to being infinite-dimensional.

\section{Conclusion}

We have identified the position and momentum of a TEM$_{00}$
optical beam. By showing that $\hat x$ and $\hat p$ are conjugate
observables that satisfy the Heisenberg commutation relation, a
continuous variable $x$-$p$ entanglement scheme is proposed. This
proposed entanglement, as considered by EPR \cite{EPR}, was
characterised using a generalised form of the {\it inseparability
criterion}.

We further explored a form of spatial entanglement which has
applications in quantum imaging. The detection schemes for quantum
imaging are typically array detectors. In this article, we
considered the split detector. We utilised the one-dimensional
spatial squeezing work of Treps {\it et al.} \cite{treps1d} and
proposed a spatial homodyning scheme for the spatial squeezed
beam. By identifying the conjugate observables for the spatial
squeezed beam as the amplitude and phase quadratures of the
flipped mode, we showed that split detection-based spatial
entanglement can be obtained.

\begin{acknowledgments}

We would like to thank Vincent~Delaubert and Hans-A.~Bachor for
fruitful discussions. This work was supported by the Australian
Research Council Centre of Excellence Programme. The Laboratoire
Kastler-Brossel of the Ecole Normale Superieure and University
Pierre et Marie Curie is associated via CNRS.

\end{acknowledgments}


\begin{thebibliography}{99}

\bibitem{EPR}{A.~Einstein, B.~Podolsky and N.~Rosen, Phys.~Rev. {\bf 47}, 777 (1935)}

\bibitem{silberhorn}{Ch.~Silberhorn, P.~K.~Lam, O.~Weiss, F.~K\"onig, N.~Korolkova and G.~Leuchs, Phys.~Rev.~Lett. {\bf 86}, 4267 (2001)}.

\bibitem{ou}{Z.~Y.~Ou, S.~F.~Pereira, H.~J.~Kimble and K~.C.~Peng, Phys.~Rev.~Lett. {\bf 68}, 3664 (1992).}

\bibitem{bowen-pol}{W.~P.~Bowen, N.~Treps, R.~Schnabel and P.~K.~Lam, Phys.~Rev.~Lett. {\bf 89}, 253601 (2002).}

\bibitem{howell}{J.~C.~Howell, R.~S.~Bennick, S.~J.~Bentley and R.~W.~Boyd, Phys.~Rev.~Lett. {\bf 92}, 210403 (2003).}

\bibitem{gatti4}{A.~Gatti and L.~A.~Lugiato, Phys.~Rev.~A {\bf 52}, 1675 (1995).}

\bibitem{gatti}{A.~Gatti, E.~Brambilla, L.~A.~Lugiato and M.~I.~Kolobov, Phys.~Rev.~Lett. {\bf 83}, 1763 (1999).}

\bibitem{gatti1}{A.~Gatti, E.~Brambilla, L.~A.~Lugiato and M.~I.~Kolobov, J.~Opt.~B {\bf 2}, 196 (2000).}

\bibitem{gatti2}{A.~Gatti, L.~A.~Lugiato, K.~I.~Petsas and I.~Marzoli, Europhys.~Lett. {\bf 46}, 461 (1999).}

\bibitem{gatti3}{A.~Gatti, E.~Brambilla and L.~A.~Lugiato, Phys.~Rev.~Lett. {\bf 90}, 133603 (2003).}

\bibitem{lugiato}{L.~A.~Lugiato, A.~Gatti and E.~Brambilla, J.~Opt.~B {\bf 4}, S176 (2002).}

\bibitem{abouraddy}{A.~F.~Abouraddy, B.~E.~A.~Saleh, A.~V.~Sergienko and M.~C.~Teich, Phys.~Rev.~Lett. {\bf 87}, 123602 (2001).}

\bibitem{pittman}{T.~B.~Pittman, Y.~H.~Shih, D.~V.~Strekalov and A.~V.~Sergienko, Phys.~Rev.~A {\bf 52}, R3429 (1995).}

\bibitem{abouraddy1}{A.~F.~Abouraddy, B.~E.~A.~Saleh, A.~V.~Sergienko and M.~C.~Teich, Opt.~Exp. {\bf 9}, 498 (2001).}

\bibitem{sokolov}{I~.V.~Sokolov, M.~I.~Kolobov, A.~Gatti and L.~A.~Lugiato, Opt.~Comm. {\bf 193}, 175 (2001).}

\bibitem{sokolov1}{I.~V.~Sokolov, J.~Opt.~B {\bf 2}, 179 (2000).}

\bibitem{hsu}{M.~T.~L.~Hsu, V.~Delaubert, P.~K.~Lam and W.~P.~Bowen, J.~Opt.~B {\bf 6}, 495 (2004).}

\bibitem{griffiths}{D.~J.~Griffiths, Introduction to Quantum Mechanics, Prentice-Hall Inc., New Jersey (1995).}

\bibitem{treps1d}{N.~Treps, U.~Andersen, B.~Buchler, P.~K.~Lam, A.~Ma\^\i tre, H.-A.~Bachor and C.~Fabre,  Phys.~Rev.~Lett. {\bf 88}, 203601 (2002).}

\bibitem{treps2d}{N.~Treps, N.~Grosse, W.~P.~Bowen, C.~Fabre, H.-A.~Bachor and P.~K.~Lam, Science {\bf 301}, 940 (2003).}

\bibitem{treps2dlong}{N.~Treps, N.~Grosse, W.~P.~Bowen, M.~T.~L. Hsu, A.~Ma\^\i tre, C.~Fabre, H.-A.~Bachor and P.~K.~Lam, J.~Opt.~B {\bf 6}, S664 (2004).}

\bibitem{duan}{L.-M.~Duan, G.~Giedke, J.~I.~Cirac and P.~Zoller, Phys.~Rev.~Lett. {\bf 84}, 2722 (2000)}

\bibitem{reid}{M.~D.~Reid and P.~D.~Drummond, Phys.~Rev.~Lett. {\bf 60}, 2731 (1988)}

\bibitem{fabre}{C.~Fabre, J.~B.~Fouet and A.~Ma\^\i  tre, Opt.~Lett., {\bf 76}, 76 (2000).}

\bibitem{Kolobov} M.~I.~Kolobov, Rev. Mod. Phys, {\bf 71}, 1539 (1999)

\bibitem{trepsmulti}{N.~Treps, V.~Delaubert, A.~Ma\^\i tre, J.~M.~Courty and C.~Fabre, {\bf quant-ph/0407246}.}

\bibitem{bowen-pol1}{W.~P.~Bowen, N.~Treps, R.~Schnabel and P.~K.~Lam, Phys.~Rev.~Lett. {\bf 89}, 253601 (2002).}

\bibitem{bowen-epr1}{W.~P.~Bowen, R.~Schnabel, P.~K.~Lam and T.~C.~Ralph, Phys.~Rev.~Lett. {\bf 90}, 043601 (2003).}

\bibitem{bowen-epr}{W.~P.~Bowen, R.~Schnabel, P.~K.~Lam and T.~C.~Ralph, Phys.~Rev.~A {\bf 69}, 012304 (2004).}

\bibitem{tan}{S.~Tan, Phys.~Rev.~A {\bf 60}, 2752 (1999).}

\bibitem{mancini}{S.~Mancini, V.~Giovannetti, D.~Vitali and P.~Tombesi, Phys.~Rev.~Lett. {\bf 88}, 120401 (2002).}

\end{thebibliography}
\end{document}